# Buried and ridge graphene-based silicon waveguides for broadband polarization-insensitive amplitude and phase modulators

Mohamad Najafi Hajivar and Mahmood Hosseini-Farzad*
*Department of Physics, College of Sciences, Shiraz University, Shiraz, Iran*
*\*hosseinif@shirazu.ac.ir*

**Abstract:** In this paper, four easy-to-fabricate graphene-based Si waveguide modulators are presented to overcome the strong polarization dependency of graphene-based modulators. The modulation features of two newly proposed structures, i.e. two graphene-based buried silicon waveguides in addition to two standard ridge silicon waveguides at the telecommunication wavelength of $\lambda = 1.55 \, \mu m$ are studied. The results show that for certain widths of each waveguide (the height is constant), the amplitude and phase modulations clearly become polarization-insensitive. The amplitude modulation depths for both the TE and TM modes are equal for these optimized waveguides with the precision of $10^{-4} \, dB/\mu m$. Moreover, in some of the proposed modulators, the maximum variations of the real parts of the effective mode indices (EMI) for both the TE and TM modes coincide with each other with an excellent precision ($5 \times 10^{-5}$). This precision value is much smaller than the standard criterion value for confirming a polarization-insensitive phase modulation. Furthermore, the performances of all the structures are studied for all optical telecommunication wavelengths. Even without making any changes to the structures designed at $1.55 \, \mu m$ at appropriate wavelength intervals, the structures exhibit polarization-insensitive behaviors.

## 1. Introduction

Due to the ever-increasing need of human beings for large data transfers, the need for faster, smaller, and less power-consuming electro-optical modulators has become apparent [1]. Therefore, traditional modulators constructed based on the Kerr, Pockels, Franz-Keldysh, and plasma-dispersion effects will probably not be able to meet the above-mentioned needs any more [2]. In addition, these modulators (such as silicon and lithium niobate ones) have a large footprint (on the order of millimeters to even a few centimeters) and require a high electrical power to convert electrical signals to optical ones [1, 2]. Furthermore, in the phase modulation, controlling the polarization of the incident light has a crucial effect on the performance of the device. These deficiencies have prompted researchers to utilize new 2D materials (such as graphene which has striking optical properties) in sub-micron photonic structures in order to design and fabricate high-performance electro-optical modulators [1, 3]. Researchers have proposed many methods (such as electrode sets [4] and bidirectional modulation [5]) for constructing phase modulators with polarization-insensitive properties. However, the need for a device with a long waveguide, a large operating voltage, and a small bandwidth is the major limitation of these methods. Therefore, polarization-insensitive modulators (especially the phase types) with compact dimensions, low insertion loss, and high bandwidth are very useful. Accordingly, the first graphene-based electro-optical modulator with the mode power attenuation (MPA) of about 0.1 dB⁄μm and the length of 40 μm was fabricated by Liu *et al.* [6]. A little later, an electro-absorption modulator based on a double-layer graphene silicon waveguide with the MPA of about 0.16 dB⁄μm was reported by the same group [7]. Thereafter, many graphene-based electro-absorption and phase modulators have been designed and fabricated in various photonic structures such as slab waveguides [8–16], graphene-covered fibers [17–19], Mach-Zehnder interferometer (MZI) phase modulators [20–23], ring resonators [24–27], plasmonic and hybrid modes [28–32], and slot waveguides [33–36]. While adding graphene to modulators has proved to be very useful, standard graphene modulators based on slab waveguides have a major disadvantage: the dependence of their optical loss on the mode polarization [37]. Recently, many efforts have been made to propose structures for eliminating the sensitivity of graphene modulators to polarization [38–44]. Although their results are desirable, some of the proposed structures have some complexities in terms of fabrication and have a lot of stages. In addition, these efforts have focused on either a phase or an amplitude modulator.

In this paper, four new and easy-to-fabricate structures have been introduced for high-performance polarization-insensitive graphene-based electro-optical modulators for both amplitude and phase modulations. Furthermore, some of the proposed structures operate very well in a wide range of wavelengths as they do at the wavelength of 1.55 μm. The dimensions of the proposed structures, in the ridge and buried shapes, are adjusted to achieve the same performance for the TM and TE polarization modes. The results show that for each structure there are two widths in which the modulator is not sensitive to polarization at the wavelength of 1.55 μm. Moreover, the effective mode indices (EMI) have been calculated numerically for all the structures with the adjusted dimensions in terms of the chemical potential of graphene. Besides, its real and imaginary parts (MPA is used instead of the imaginary part) have been studied in detail. Conventionally, for amplitude modulation, the variation of MPA (i.e. the difference in its value between the on- and off-states also called 'modulation depth') is a key parameter for the evaluation of efficiency. The results indicate that the MPA curves of the TM and TE modes coincide with each other for all the proposed optimized structures showing the polarization-independent amplitude modulation. Furthermore, with the obtained values for MPA in the off-state (the smallest value is about 0.3 dB/μm), short footprint modulators can be fabricated based on the proposed structures so that only a length of 10 μm is sufficient for a 3-dB modulation. In the on-state, the value of MPA is very small (0.01 dB/μm).

The current study shows that the proposed structures with the adjusted dimensions can also act as polarization-insensitive phase modulators. Due to the direct relationship between the real part of EMI and the phase shifting of the propagating mode in the waveguide, the real part of EMI has been studied in terms of the chemical potential of graphene layers. The results for the ridge and buried waveguides show that the fundamental TE and TM modes have the same behavior. For each structure, the corresponding curves for the TE and TM modes coincide with each other with a precision of $10^{-4}$ and even less. The maximum variation of the real part of EMI with respect to the chemical potential is nearly equal to 0.02 which is more than the values reported before [20]. Moreover, this value demonstrates that the required mode propagating length for a π-phase shift is nearly 40 μm. Finally, the robustness of the presented polarization-insensitive modulators when they are used at other operational wavelengths in the optical telecommunication region is also studied.

## 2. The tunable optical properties of graphene

Graphene is one of the thinnest materials artificially synthesized in professional laboratories. It consists of a 2D arrangement of carbon atoms each of which has three covalent bonds with its neighbors in a honeycomb lattice. The fourth electron stays free which is the origin of the extraordinary properties of graphene [45]. In spite of the high in-plane mobility of the free electrons, they are bounded in the out-of-plane direction. Hence, the in-plane electrical conduction and consequently the in-plane dielectric constant of graphene totally differ from the out-of-plane ones depending on the substrate material. This means that graphene is an anisotropic optical material [46]. Due to the gapless band structure of graphene, when the density of the electric charge carriers is altered, a quantum phenomenon called 'Pauli-blocking' happens. In fact, by changing the density of the carriers in graphene which is equal to changing the chemical potential (the Fermi energy), the electrons cannot absorb photons with an energy lower than twice as much as the Fermi energy. As a result, absorption suddenly drops and graphene becomes transparent to light. The in-plane conductivity (and as a result, the permittivity) of graphene has been theoretically modeled by different approaches such as the Kubo formula [46], the random phase approximation [46], and the Kramers–Kronig relations [47]. The out-of-plane permittivity of graphene is usually considered as a constant value of 2.5. In this paper, the Kramers–Kronig relations were used to define the dielectric constant of graphene. Thus, the real and imaginary parts of graphene are obtained by the following equations, respectively [47]:

$$\varepsilon'_g(\mu, E_p) = 1 + \frac{e^2}{8\pi E_p \varepsilon_0 d_g} \ln\left(\frac{(E_p+2|\mu_c|)^2 + \Gamma^2}{(E_p-2|\mu_c|)^2 + \Gamma^2}\right) - \frac{e^2}{\pi E_p^2 \varepsilon_0 d_g} \frac{|\mu_c|}{E_p^2 + (1/\tau)^2} \qquad (1)$$

$$\varepsilon''_g(\mu_c, E_p) = \frac{e^2}{4 E_p \varepsilon_0 d_g}\left[1 + \frac{1}{\pi}\left\{\tan^{-1}\left(\frac{E_p-2|\mu_c|}{\Gamma}\right) - \tan^{-1}\left(\frac{E_p+2|\mu_c|}{\Gamma}\right)\right\}\right] + \frac{e^2}{\pi \tau E_p \varepsilon_0 d_g} \frac{|\mu_c|}{E_p^2 + (1/\tau)^2} \qquad (2)$$

In the above relations, $d_g$ is the thickness of the graphene layer that is equal to $1\ nm$ in the current study. The interband transition broadening $\Gamma$ is set to $110\ meV$ from the graphene reflection spectrum. The free-carrier scattering rate $1/\tau$ can be assumed to be zero because of its negligible effect on the real and imaginary parts of the dielectric constant at the incident photon frequency $\omega$ (the near-infrared region). Therefore, the last term in Eq. (2) is ignored [47]. Besides, $e$ is the electric charge of the electron, $E_p$ is the energy of the incident photon in eV (here the operation wavelength is $1.55\ \mu m$, hence, $E_p = 0.8\ eV$), $\varepsilon_0$ is the vacuum permittivity, and $\mu_c$ is the chemical potential of graphene in $eV$ [47]. Accordingly, the in-plane dielectric constant of graphene is $\varepsilon_\parallel(\mu_c, E_p) = \varepsilon'_g(\mu_c, E_p) + i\,\varepsilon''_g(\mu_c, E_p)$. The chemical potential of graphene can be tuned by using a gate voltage [6]:

$$\mu_c = \hbar v_f \sqrt{\pi |\eta(V_g + V_0)|} \qquad (3)$$

where $\hbar$ is the reduced Planck constant, $v_f$ is the Fermi velocity which is equal to $0.9 \times 10^6\ ms^{-1}$, $\eta = 9 \times 10^{16}\ m^{-2}V^{-1}$ is estimated using a parallel-plate capacitor model of the device, $V_g$ is the gate voltage, and $V_0$ is the offset voltage caused by the natural doping of graphene [6]. Therefore, by applying appropriate changes to $V_g$, the desired amount of $\mu_c$ is obtained.

In Fig. 1, the variations of the real and imaginary parts of the dielectric constant of graphene as the functions of the chemical potential ($\mu_c$) are plotted. At a certain value of the chemical potential, graphene experiences a maximum in its real part of the dielectric constant, while its imaginary part decreases fast to zero concurrently. In fact, this value for the chemical potential is exactly equal to half of the photon energy and Pauli-blocking is responsible for this behavior [3]. Based on the above discussion, since the photon energy is 0.8 eV, the maximum value of the real part of the dielectric constant and a sudden drop in its imaginary part both occur at the value of $\mu_c = 0.4\ eV$. Therefore, the chemical potential values of less than half of the photon energy for $\mu_c$ are suitable for the off-state (high absorption), whereas the chemical potential values of more than half of the photon energy should be selected for the on-state in the amplitude modulation mechanism.

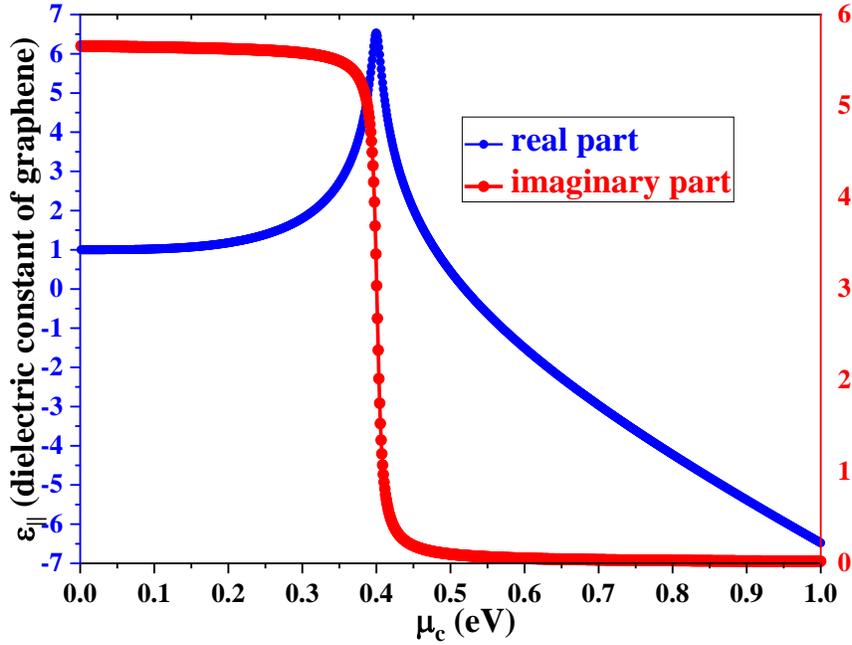

Fig. 1. The real and imaginary parts of the graphene dielectric as the functions of the chemical potential. The wavelength of the incident light is 1.55 $\mu m$ ($E_p = 0.8\ eV$) and the graphene thickness is 1 nm.

### 3. The modulator structures and the analysis method

In this paper, four structures are proposed for polarization-insensitive electro-optical modulators based on rectangular cross-section silicon waveguides with their surfaces covered by graphene layers. Two of them have a ridge geometry and the other two are buried waveguides (see Fig. 2). First, the structure composed of a Si-slab buried in a SiO$_2$ substrate with two layers of graphene at the bottom and one layer on each of the sidewalls (Fig. 2(a)) is considered. It should be noted that two horizontal graphene layers (in the x-direction) are also on the top of the SiO$_2$ substrate and are located on the left and right sides of the buried and ridge waveguides in order to facilitate the connection of the structures to the gate voltages or electrodes. The second structure is also a buried waveguide with only one graphene layer on its floor and another graphene layer on its roof (see Fig. 2(b)). The third configuration is sketched in Fig. 2(c). It consists of a Si-ridge waveguide lying on a SiO$_2$ substrate, a single layer of graphene on each sidewall, and two graphene layers on the top of the waveguide. The last structure is depicted in Fig. 2(d). It has the same structure as that of (c) but has only a single layer of graphene on its top and also has a single layer on its bottom. In all the structures, an insulating layer (e.g. Al$_2$O$_3$) is between the Si medium and the graphene layers and also between the two adjacent graphene layers in order to prevent the transfer of charge carriers between them. Increasing or decreasing the density of electric charge carriers in graphene layers in the structures can be analogized to the charging or discharging of a capacitor. The larger the capacitance, the less the power needed to charge and discharge it. It should be noted that in structures (a) and (c), two graphene layers that lie at the bottom and top of the Si core, respectively, form a planar capacitor. According to the capacitance equation of the planar capacitor, $C = \varepsilon\varepsilon_0 A/d$, the separation ($d$ which is the thickness of the Al$_2$O$_3$ spacer between the graphene layers) between the two graphene layers is sufficiently small and the capacitance is ideally large. In structures (b) and (d), the horizontal graphene layers stretch to the two sides of the waveguide cores to overlap each other. Hence, the capacitance of the capacitor is significantly enhanced.

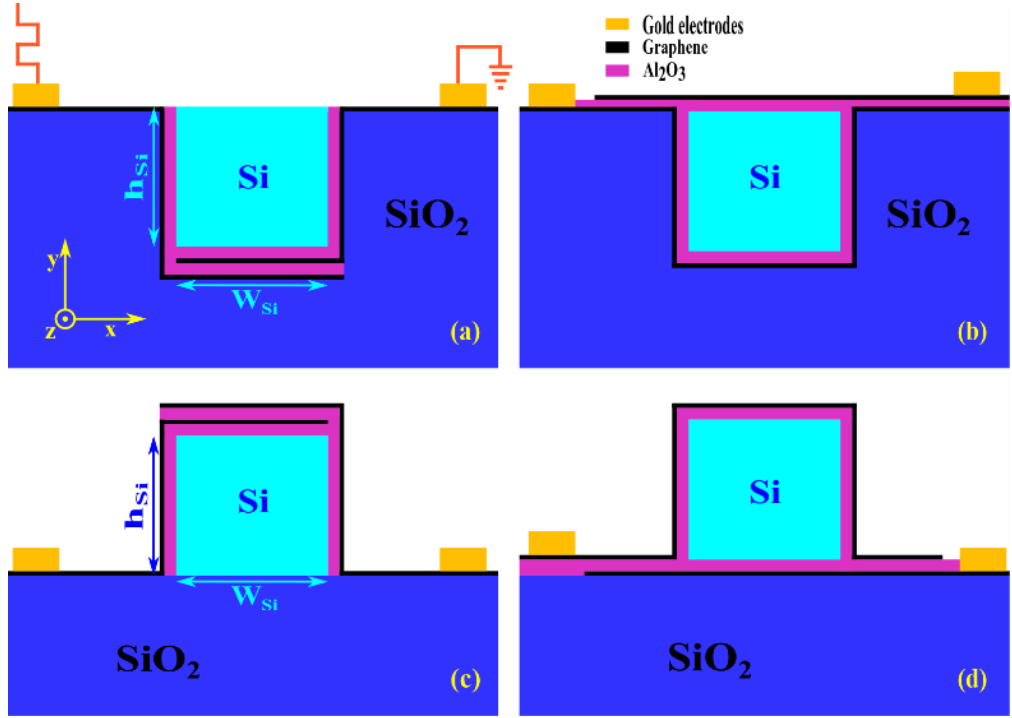

Fig. 2. The cross-section of the proposed structures: a) 3-side-graphene-covered buried waveguide with two graphene layers at the bottom and one layer on each sidewall. b) 4-side-graphene-covered buried waveguide with one layer on each side. c) 3-side-graphene-covered ridge waveguide with two graphene layers on the top and one layer on each sidewall. d) 4-side-graphene-covered ridge waveguide with one graphene layer on each side. The height of the Si waveguide ($h_{si}$) was fixed at 300 nm.

To study these structures accurately, it is necessary to consider graphene as an anisotropic material. It is also important to note that the graphene layer lies horizontal (in the x-z plane) or vertical (in the y-z plane) inside the waveguide structure. Therefore, the dielectric constant tensor of the horizontal graphene layers can be written as:

$$\varepsilon = \begin{pmatrix} \varepsilon_\parallel & 0 & 0 \\ 0 & \varepsilon_\perp & 0 \\ 0 & 0 & \varepsilon_\parallel \end{pmatrix} \quad (4)$$

In addition, the corresponding tensor for vertical graphene layers is:

$$\varepsilon = \begin{pmatrix} \varepsilon_\perp & 0 & 0 \\ 0 & \varepsilon_\parallel & 0 \\ 0 & 0 & \varepsilon_\parallel \end{pmatrix} \quad (5)$$

$\varepsilon_\parallel$ and $\varepsilon_\perp$ have already been introduced by relations $\varepsilon_\parallel = \varepsilon'_g + i\varepsilon''_g$ and $\varepsilon_\perp = 2.5$. The propagation direction in the waveguides is along the z axis. Thus, for horizontal (vertical) graphene layers, the out-of-plane direction is along the y axis (x axis). The height of the Si slab ($h_{si}$) is fixed at the value of 300 nm for all the structures. However, the width ($W_{Si}$) of the Si slab varies from 210 to 390 nm for the buried structures and from 180 to 390 nm for the ridge structures. Some properties of the fundamental TE and TM modes in the proposed structures (such as MPA at $\mu_c = 0\ (eV)$) are presented in different parts of Fig. 3. According to this figure, the MPA for the TM mode has fewer changes with respect to the variations of the waveguide width compared to that of the TE mode. As shown in Fig. 3, the TE mode curves have an obvious maximum in MPA. The curves for two ridge structures have more drastic variations compared to those of the buried structures. In addition, the curves of the TM and TE modes cross each other at two points. These intersection points are much desired since at these points the MPAs for the two different polarization modes become equal. Therefore, these points demonstrate the optimized dimensions of the waveguide to overcome the functionality dependence of graphene modulators on the incident light polarization.

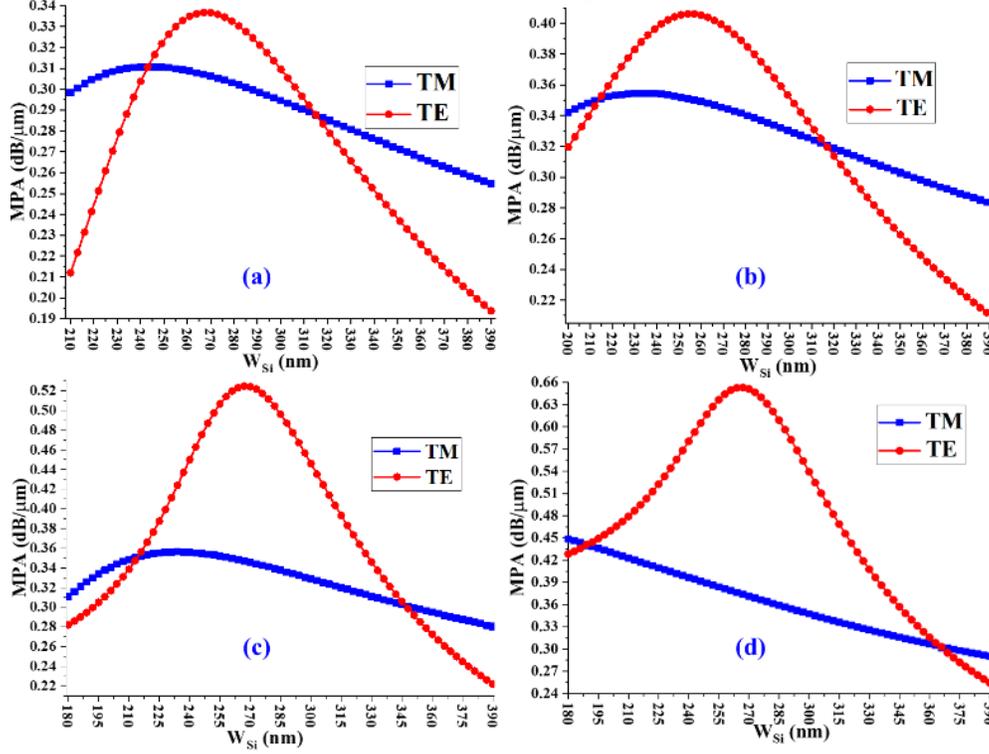

Fig. 3. MPAs ($dB/\mu m$) for two different polarization modes (TE & TM) in terms of $W_{Si}$ for a) 3-side-graphene-covered buried waveguide (Fig. 2-a) b) 4-side-graphene-covered buried waveguide (Fig. 2-b) c) 3-side-graphene-covered ridge waveguide (Fig. 2-c) d) 4-side-graphene-covered ridge waveguide (Fig. 2-d)

In Fig. 4, the differences between the real parts of the EMI (for the chemical potential values of $\mu_c = E_p/2 = 0.4\ eV$ and $1.0\ eV$) as a function of the waveguide width ($W_{Si}$) are calculated for the TE ($\Delta n_{eff}^{TE}$) and TM ($\Delta n_{eff}^{TM}$) modes. These chemical potential values are selected according to Fig. 1 in order to specify the maximum alteration in the real parts of EMI in the two modes. By comparing the different parts of Fig. 3 with their corresponding counterparts in Fig. 4, a similar behavior for the variations of $\Delta n_{eff}$ and MPA with respect to $W_{Si}$ can be seen. Besides, the intersection points of the TE and TM curves for the same structures in the two figures are identical.

Thus, based on this comparison, a very interesting result emerges, i.e. the proposed structures are very good candidates for polarization-insensitive modulators (both amplitude and phase types). In other words, by selecting an appropriate size for the graphene-based Si waveguide, the structure can act not only as a polarization-insensitive phase modulator but also as a polarization-insensitive amplitude modulator. The exact values of $W_{Si}$ at the first and second intersection points obtained from Figs. 3 and 4 along with the corresponding MPAs and EMIs at these waveguide widths are presented in Table 1.

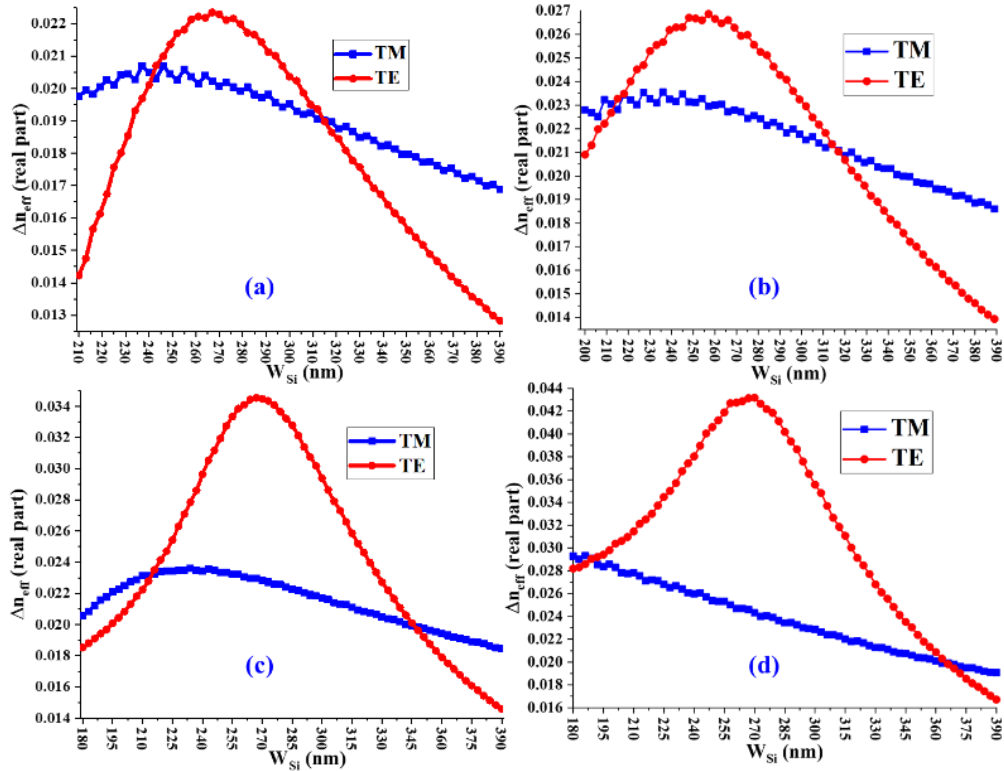

Fig. 4. The difference between the real part of the EMI in $\mu_c = 0.4\ (eV)$ and $\mu_c = 1\ (eV)$ in terms of $W_{Si}$ for a) 3-side-graphene-covered buried waveguide (Fig. 2-a) b) 4-side-graphene-covered buried waveguide (Fig. 2-b) c) 3-side-graphene-covered ridge waveguide (Fig. 2-c) d) 4-side-graphene-covered ridge waveguide (Fig. 2-d)

The curves in Fig. 4 have some ripples whose origin is the special properties of graphene. Therefore, the curves are not as smooth as those in the previous figure (MPAs at $\mu_c = 0\ eV$). According to Fig. 1, when the chemical potential is increased, the real part of the dielectric constant of graphene begins to increase until it experiences a cusp-like maximum at $\mu_c = E_p/2$. Then it decreases to zero and eventually becomes negative (for instance at $\mu_c = 1\ eV$). The negative values of the dielectric constant provide a proper opportunity for the excitation of some surface modes. Therefore, unlike ordinary noble metals such as Au and Ag, both TM and TE surface modes can be propagated along the graphene layers when the real parts of the dielectric constant of the graphene layers become negative [48]. The surface modes affect the EMIs of the propagating modes of the structures and are responsible for the small ripples in the curves of Fig. 4 and those of Fig. 13. In addition, the difference of the real part of EMI for two chemical potential values at $\mu_c = 0.4\ eV$ and $\mu_c = 0.5\ eV$ (at which the dielectric constant of graphene is not negative) was calculated. Hence, the surface modes no longer existed and the ripples disappeared completely. Afterward, the numerical simulations were repeated for the actual thickness of graphene layers ($d_g = 0.34\ nm$) and the surface modes and the ripples were still observed. Fig. 5 shows the electric field distributions for the dominant (normal) component of the TM and TE modes typically in structure (d) (with $W_{Si} = 365\ nm$) when $\mu_c = 1\ eV$. By magnifying the field distributions on the graphene layers, (the y-component for the TM mode and the x-component for the TE mode), some surface modes similar to surface plasmons are clearly observed.

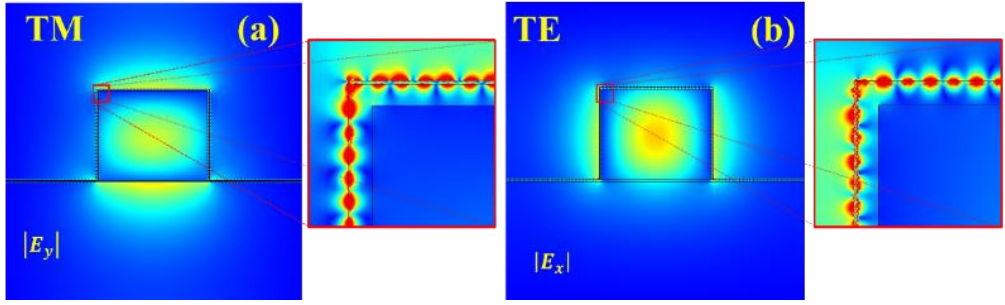

*Fig. 5. The normal component of the electric field distributions for the fundamental a) TM and b) TE modes of structure (d) with $W_{Si} = 365$ nm and $\mu_c = 1$ eV at $\lambda = 1.55$ μm. The insets in each figure magnify the details of the field distributions belonging to the graphene surface modes.*

Table 1. The calculated features of the proposed structures with the best $W_{Si}$ for the fundamental TE and TM modes

| Structure | Intersection point | Best $W_{Si}$ (nm) | $MPA_{TE}$ $(dB/\mu m)$ | $MPA_{TM}$ $(dB/\mu m)$ | $\Delta n_{eff}^{TE}$ | $\Delta n_{eff}^{TM}$ | $\|\Delta n_{eff}^{TE} - \Delta n_{eff}^{TM}\|$ |
|---|---|---|---|---|---|---|---|
| a | First | 242 | 0.31210 | 0.31206 | 0.02058 | 0.02066 | $8 \times 10^{-5}$ |
| a | Second | 315 | 0.28953 | 0.28875 | 0.01884 | 0.01898 | $1.4 \times 10^{-4}$ |
| b | First | 214 | 0.35067 | 0.35064 | 0.02320 | 0.02305 | $1.5 \times 10^{-4}$ |
| b | Second | 316 | 0.32131 | 0.32118 | 0.02112 | 0.02117 | $5 \times 10^{-5}$ |
| c | First | 214 | 0.35100 | 0.35017 | 0.02302 | 0.02317 | $1.5 \times 10^{-4}$ |
| c | Second | 346 | 0.30347 | 0.30221 | 0.01989 | 0.01994 | $5 \times 10^{-5}$ |
| d | First | 190 | 0.44128 | 0.44024 | 0.02917 | 0.02900 | $1.7 \times 10^{-4}$ |
| d | Second | 365 | 0.30508 | 0.30497 | 0.02007 | 0.02001 | $6 \times 10^{-5}$ |

## 4. Electro-absorption modulation

In this section, the amplitude modulations for the TE and TM fundamental modes of the different proposed structures with optimized $W_{Si}$ (Table 1) values are studied. In addition, the electric field distributions for each TE and TM mode at the wavelength of 1.55 μm for these optimized buried and ridge graphene-based silicon waveguides are investigated. As stated above, when the Fermi energy or the chemical potential of graphene increases and passes a cut-off point $(\mu_c = E_p/2)$, the electrons in the valence band are no longer able to absorb the incident photons (with energy $E_p$). Thus, graphene becomes transparent to light. Therefore, by tuning the chemical potential, the imaginary part of EMI of the whole structure and subsequently the absorption of graphene can be modulated. Here, the MPA of the structure is directly related to the imaginary part of EMI:

$$MPA = 40\pi \frac{n_I}{\lambda} \log(e) \qquad (6)$$

Here $n_I$ is the imaginary part of the EMI of the structure and $\lambda$ is the wavelength of light. The MPAs of the TE and TM modes as the functions of the chemical potential for two types of buried waveguides (parts (a) and (b) in Fig. 2) with optimized $W_{Si}$ values (extracted from Fig. 3) are plotted in Fig. 6. As shown, the general behavior of MPA in the proposed waveguides is the same as the imaginary part of the dielectric constant of graphene with respect to its chemical potential. In addition, as can be seen in the different parts of Fig. 6, the values of MPA for the TE mode coincide with those of the TM mode with the precision of less than $10^{-3}\ dB/\mu m$. This is in fact an interesting result and the proposed goals of the current study for the polarization independence of amplitude modulation in these waveguides are well fulfilled.

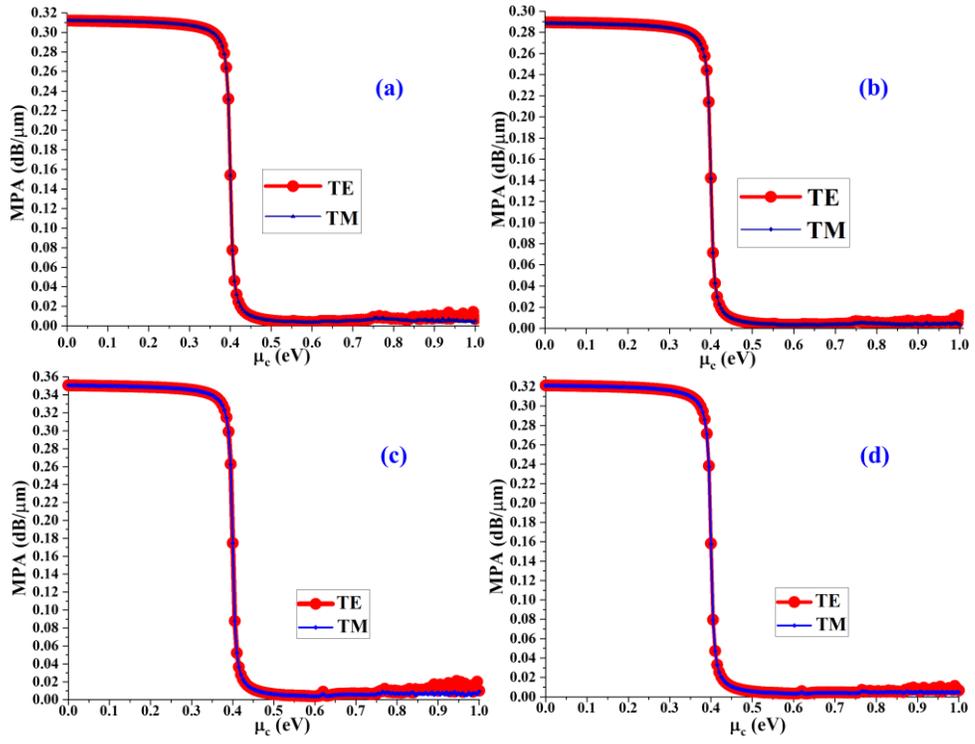

Fig. 6. The calculated MPAs for the buried structures a) 3-side-graphene-covered buried waveguide with $W_{Si} = 242\ nm$ b) 3-side-graphene-covered buried waveguide with $W_{Si} = 315\ nm$ c) 4-side-graphene-covered buried waveguide with $W_{Si} = 214\ nm$ d) 4-side-graphene-covered buried waveguide with $W_{Si} = 316\ nm$

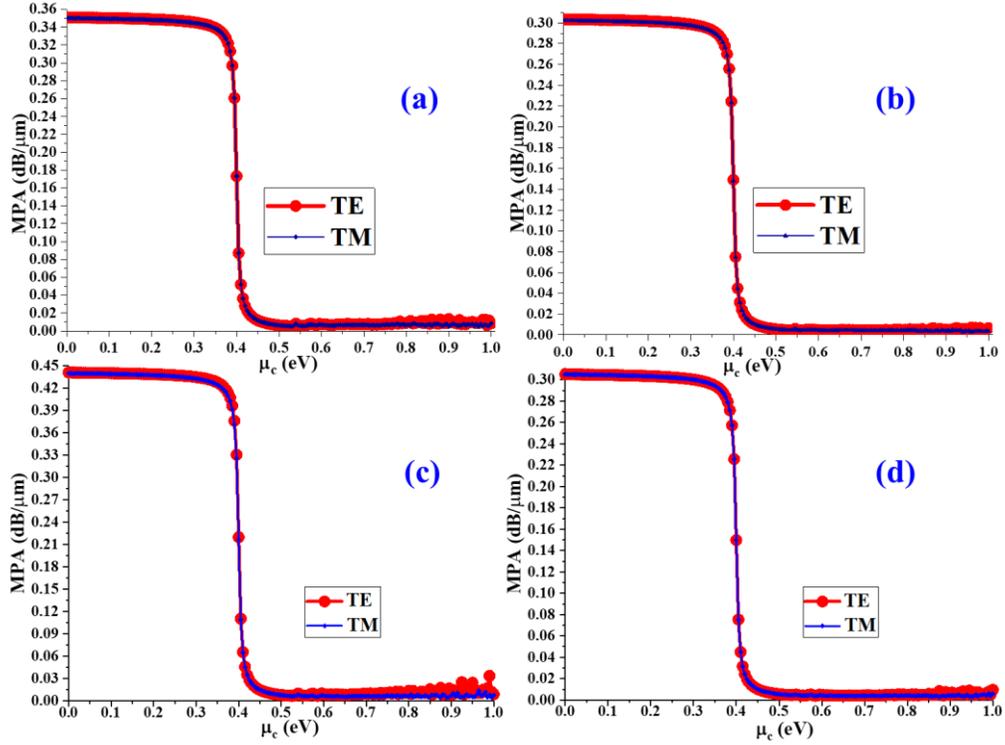

Fig. 7. The calculated MPAs for the ridge structures for a) 3-side-graphene-covered ridge waveguide with $W_{Si} = 214\ nm$ b) 3-side-graphene-covered ridge waveguide with $W_{Si} = 346\ nm$ c) 4-side-graphene-covered ridge waveguide with $W_{Si} = 190\ nm$ d) 4-side-graphene-covered ridge waveguide with $W_{Si} = 365\ nm$

As shown in Fig. 7 (a-d), the MPA values of two different polarization modes for two optimized ridge structures behave like a step function. At first, they are approximately constant but suddenly fall by passing a certain point ($\mu = 0.4\ eV$) and then become constant again with a very small value. Furthermore, the curves of the TE and TM modes have a good conformity with each other which shows that the proposed ridge structures can play their roles as polarization-insensitive electro-absorption modulators very well. By changing the chemical potential of graphene, the imaginary part of the refractive index changes and consequently the absorption coefficient or equivalently MPA changes. In fact, the modulation depth can be determined by the MPA discrepancy between the on-state and the off-state. A chemical potential value of the graphene layers must be considered as the off-state in which the MPA is maximum and consequently the transmission is minimum. On the other hand, a chemical potential value must be considered as the on-state in which the MPA is minimum and therefore the transmission is maximum. What is more, due to the relationship between the chemical potential and the applied gate voltage, in order to reduce power consumption as much as possible, the two mentioned values for the chemical potentials should be as close to each other as possible. A reasonable choice may be the two chemical potential values just before and after the sudden decrease point ($\mu_c = 0.4\ eV$). In addition, the ridge structures have a larger attenuation compared to their buried counterparts for small chemical potential values.

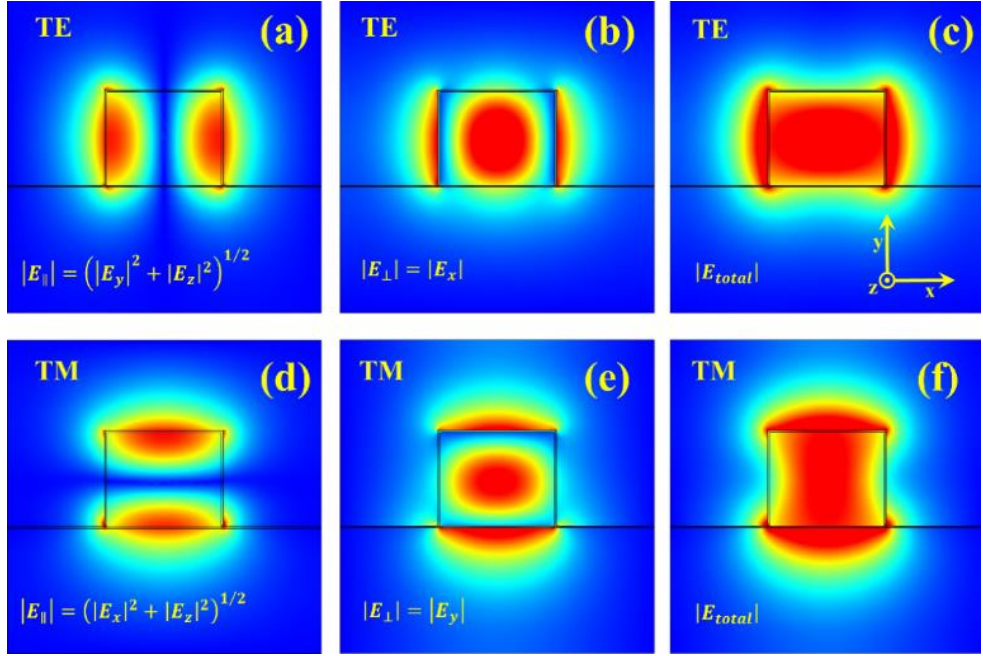

Fig. 8. a) The tangential component of the electric field of the TE mode, b) the normal component of the electric field ($|E_x|$) of the TE mode, c) the total electric field of the TE mode d) the tangential component of the electric field of the TM mode, e) the normal component of the electric field ($|E_y|$) of the TM mode and f) the total electric field of the TM mode for the 4-side-graphene-covered ridge waveguide with $W_{Si} = 354\ nm$. The scales of all the contours are the same.

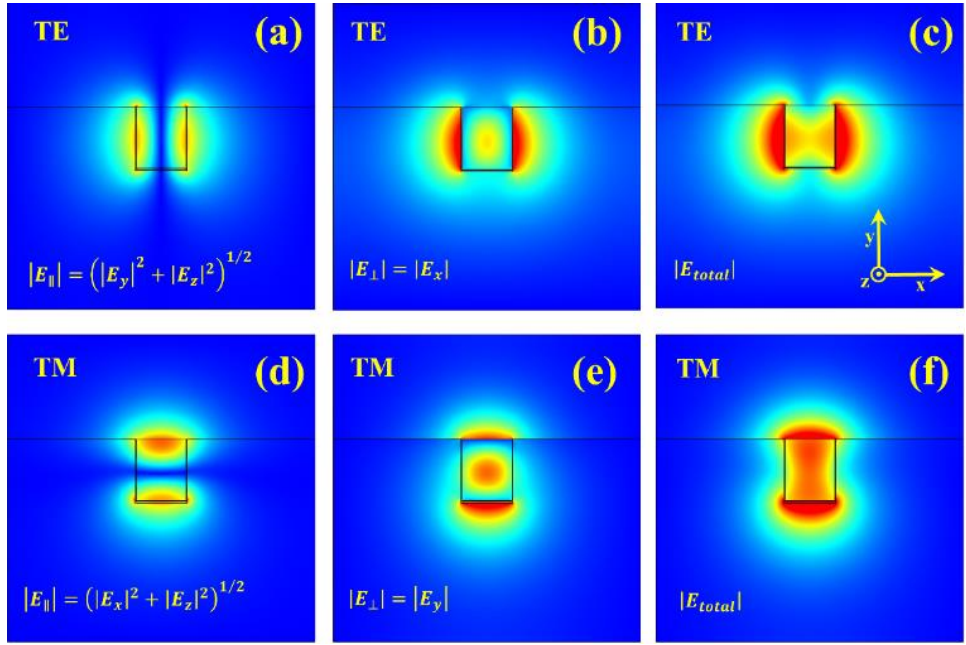

Fig. 9. a) The tangential component of the electric field of the TE mode, b) the normal component of the electric field ($|E_x|$) of the TE mode c) the total electric field of the TE mode d) the tangential component of the electric field of the TM mode, e) the normal component of the electric field ($|E_y|$) of the TM mode and f) the total electric field of the TM mode for the 3-side-graphene-covered buried waveguide with $W_{Si} = 242\ nm$. The scales of all the graphs are the same.

Here, the electric field distribution for the TE and TM fundamental modes in two of these modulators with optimized $W_{Si}$ is investigated. The z axis is the propagation direction which is parallel to all the graphene layers. At first, the 4-side-graphene-covered ridge waveguide (see part (d) of Fig. 2) with $W_{Si} = 354$ nm is investigated. Fig. 8 demonstrates these field distributions for the TE (a-c) and TM (d-f) modes. In each mode, the total electric field has two components: the field in the plane of graphene layers (tangential component) and the field perpendicular to it (normal component). For the TE mode, the distribution of these components is illustrated in parts (a) and (b) of Fig. 8, respectively, while the magnitude of the total electric field is shown in part (c) of Fig. 8. Here, the magnitude of the tangential component of the electric field is $|E_\parallel| = (|E_y|^2 + |E_z|^2)^{1/2}$ and its normal component is $|E_x|$. In the TE mode, the x-component of the electric field is considered as the normal component since it is vertical to the left and right walls of the Si waveguide. This component has very little contribution to the TE modulation process since it is normal to the graphene layers on the left and right walls of the waveguide. Moreover, $E_x$ has small in-plane values on the top and bottom walls (around their middle regions) which are also covered by graphene layers (see Fig. 8(b)). Nearly the same procedure is valid for the TM mode in this type of waveguide. However, in this case, the y-component of the electric field plays the role of the normal component (Fig. 8 (e)) and the tangential component becomes $|E_\parallel| = (|E_x|^2 + |E_z|^2)^{1/2}$ (Fig. 8(d)). Like the TE mode, the normal component of the electric field ($E_y$) in the TM mode has little contribution to the modulation task compared to the tangential one ($E_\parallel$) since it has a small amount on the graphene layers on the left and right walls of the Si waveguide (compare parts (d) and (e) of Fig. 8). Secondly, as shown in Fig. 9, as another example, the electric field distribution of the TE and TM fundamental modes is calculated for the first intersection point in an optimized buried waveguide structure depicted in Fig. 2 (a). Like Fig. 8, the tangential and normal components of the electric fields of the TE and TM modes are plotted for the 3-side-graphene-covered buried waveguide with $W_{Si} = 242\ nm$ (see Fig. 9).

## 5. Phase modulation

Here the phase modulation behavior of these graphene-based waveguides and their polarization independence in phase modulation is analyzed. For this goal, the alterations of the real parts of EMI belonging to the fundamental TE and TM modes in all the optimized waveguide structures are investigated in terms of chemical potential. The first and second rows in Fig. 10 demonstrate the variations of the real parts of the EMIs versus the chemical potential for the 3-side- and 4-side-graphene-covered buried waveguides, respectively. Similarly, the corresponding diagrams for the 3-side- and 4-side-graphene-covered ridge waveguides have been shown in the first and second rows of Fig. 11. The general behavior of the curves of the TE and TM modes which overlap each other with a good precision (Figs. 10 and 11) is similar to that of the real part of the dielectric constant of graphene versus chemical potential (Fig. 1). The differences between the real parts of EMI at its maximum ($\mu_c = 0.4\ eV$) and minimum ($\mu_c = 1.0\ eV$) for all the waveguides have been extracted from Figs. 10 and 11 and listed separately for each structure in Table 1 for the TE and TM modes. In Table 1, the values of these differences for all the waveguides are shown for the TE and TM modes by $\Delta n_{eff}^{TE}$ and $\Delta n_{eff}^{TM}$, respectively. On the other hand, the value of the quantity $\left|\Delta n_{eff}^{TE} - \Delta n_{eff}^{TM}\right|$ is a good criterion for checking the polarization-insensitive feature of phase modulation. The standard value of this quantity that confirms the existence of this feature is $5 \times 10^{-4}$ [41]. Accordingly, the results for this parameter, shown in the last column of Table 1, belong to the proposed phase modulators because they are all smaller than the permitted standard value. This demonstrates that the proposed graphene-based Si waveguides previously introduced as polarization-independent amplitude modulators can also act as good polarization-insensitive phase modulators.

Another important parameter in phase modulators is the length (here denoted by $l_\pi$) that the propagating light must travel in order to obtain a $\pi$-phase shift. It is easily calculated by the following relation:

$$\frac{2\pi \Delta n}{\lambda} l_\pi = \pi \tag{13}$$

For the modulators in the current study $\Delta n \approx 0.02$. Thus, an average value of $l_\pi$ is almost 40 $\mu m$ at the wavelength of 1.55 $\mu m$. To the best of the authors' knowledge, this value for $l_\pi$ is better than those for all the other graphene-based waveguide phase modulators reported before.

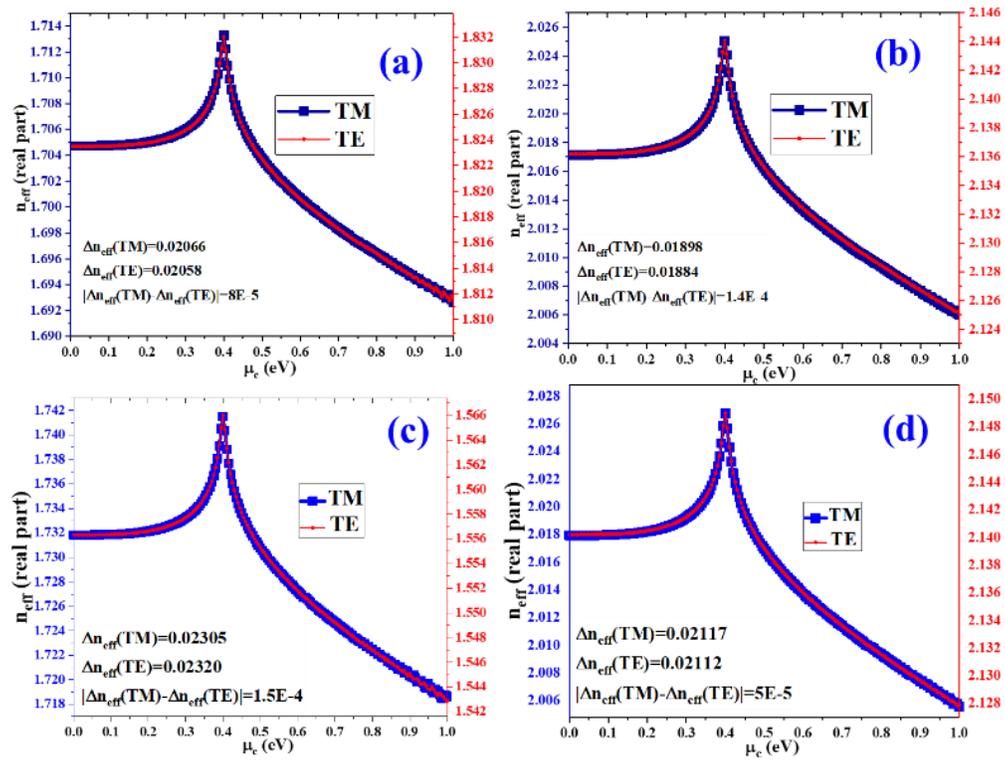

Fig. 10. The real parts of EMIs for the buried structures in terms of the chemical potential for a) the 3-side-graphene-covered buried waveguide with $W_{Si} = 242\ nm$ b) the 3-side-graphene-covered buried waveguide with $W_{Si} = 315\ nm$ c) the 4-side-graphene-covered buried waveguide with $W_{Si} = 214\ nm$, and d) the 4-side-graphene-covered buried waveguide with $W_{Si} = 316\ nm$

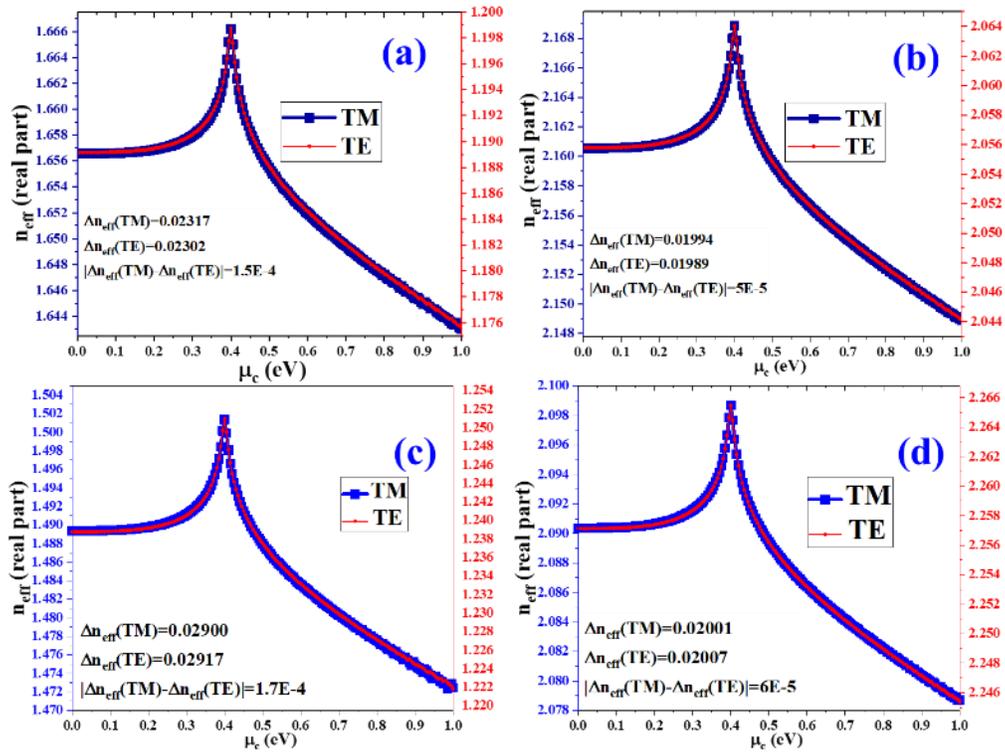

Fig. 11. The real parts of EMIs for the ridge structures as the functions of the chemical potential for a) the 3-side-graphene-covered ridge waveguide with $W_{Si} = 214$ nm b) the 3-side-graphene-covered ridge waveguide with $W_{Si} = 346$ nm c) the 4-side-graphene-covered ridge waveguide with $W_{Si} = 190$ nm, and d) the 4-side-graphene-covered ridge waveguide with $W_{Si} = 365$ nm

## 6. The study of the proposed structures in optical telecommunication wavelengths

Although the structures have been designed for polarization-insensitive amplitude and phase modulations at the wavelength of 1.55 $\mu m$, their behaviors in the optical telecommunication range are also investigated. The performances of all the structures using MPA and $\Delta n_{eff}$ as the functions of wavelength are studied. For amplitude modulation, the MPA spectra of the fundamental TE and TM modes have been calculated at $\mu_c = 0\ eV$ for all the structures with optimized $W_{Si}$ belonging to the second intersection point (Table 1). As shown in Fig. 11, the MPAs for two fundamental modes of each structure and also their difference ($\Delta MPA = |MPA_{TM} - MPA_{TE}|$) as the functions of wavelength are calculated for a better understanding of the modulation bandwidth of each optimized structure. The general features of $\Delta MPA$ spectra for all the structures are similar to each other and as already expected they experience a minimum nearly at the wavelength of 1.55 $\mu m$.

A 1 $dB$ difference in the output intensity of the TE and TM modes is an acceptable criterion showing that the polarization-insensitive feature is conserved [41]. Therefore, when a modulation length of 50 $\mu m$ is selected, the value of $\Delta MPA$ is equal to 0.02 $dB/\mu m$. Hence, the modulation bandwidth of the structures can be defined as a range of wavelengths in which $\Delta MPA$ is less than 0.02 $dB/\mu m$. Therefore, based on this criterion, structure (a) can act as a polarization-insensitive electro-absorption modulator in the wavelength range of 1300 to 1620 nm. This range for structures (b), (c), and (d) is 1300 to 1610 nm, 1420 to 1600 nm, and 1385 nm to 1600 nm, respectively (see Fig. 12 (a-d)).

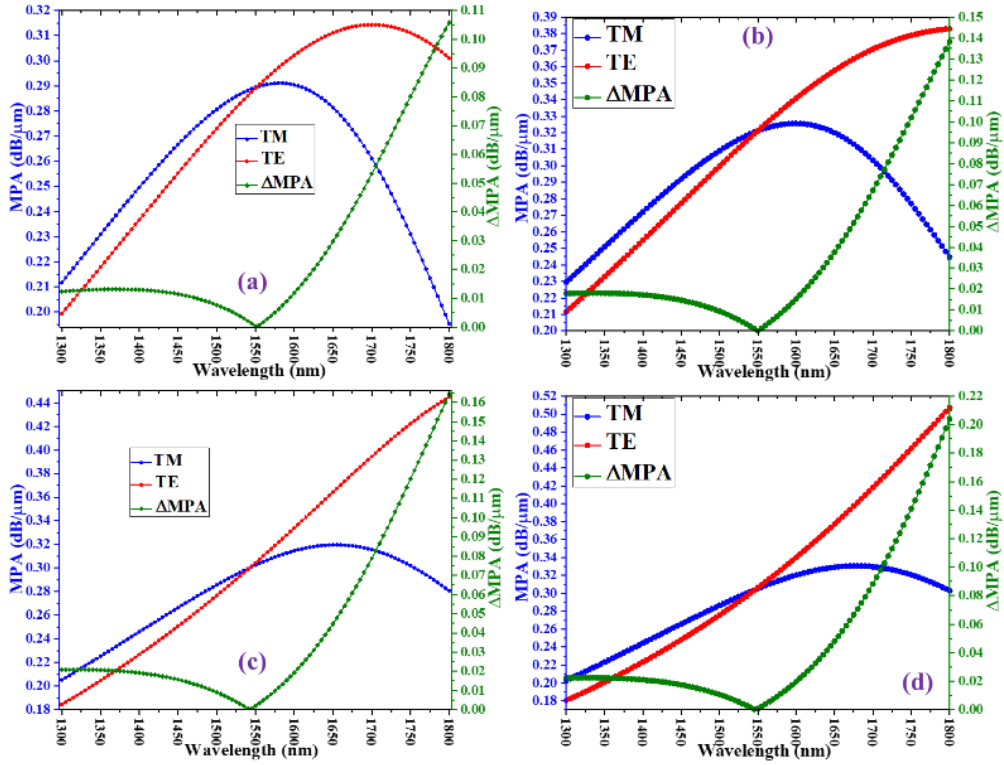

Fig. 12. The MPAs at $\mu_c = 0\ eV$ as the functions of wavelength for the waveguides with optimized $W_{Si}$ at the second intersection point (Table 1) for a) 3-side-graphene-covered buried waveguide b) 4-side-graphene-covered buried waveguide c) 3-side-graphene-covered ridge waveguide and d) 4-side-graphene-covered ridge waveguide

The same procedure can be applied for the bandwidth analysis of these optimized waveguides in phase modulation. Thus, the discrepancy between the real parts of the EMIs $(\Delta n_{eff})$ at two chemical potential values of the of graphene layers is calculated as a function of wavelength. Since the maximum value of the real parts of the EMIs occurs at the chemical potential of $\mu_c = E_p/2$ (see Figs. (9) and (10)), $\Delta n_{eff}$ has been set to the discrepancy between the real parts of EMIs at $\mu_c = E_p/2$ and $\mu_c = 1\ eV$. Moreover, the difference between the two obtained values of $\Delta n_{eff}$ for two polarization modes $|\Delta n_{eff}^{TM} - \Delta n_{eff}^{TE}|$ has also been calculated as a function of the wavelength in Fig. 12. For the polarization-insensitive phase modulation, the last parameter must be less than $5 \times 10^{-4}$[41]. Therefore, according to Fig. 12, a range of wavelengths are determined for each structure at which polarization sensitivity is negligible. Hence, by complying with this criterion and looking at various parts of Fig. 13, one can calculate the modulation bandwidths of structures (a), (b), (c), and (d) as 1500-1580 nm, 1500-1570 nm, 1510-1570 nm, and 1520-1570 nm, respectively. It is also worth noting that the optical dispersions of all the materials are considered in this wavelength study.

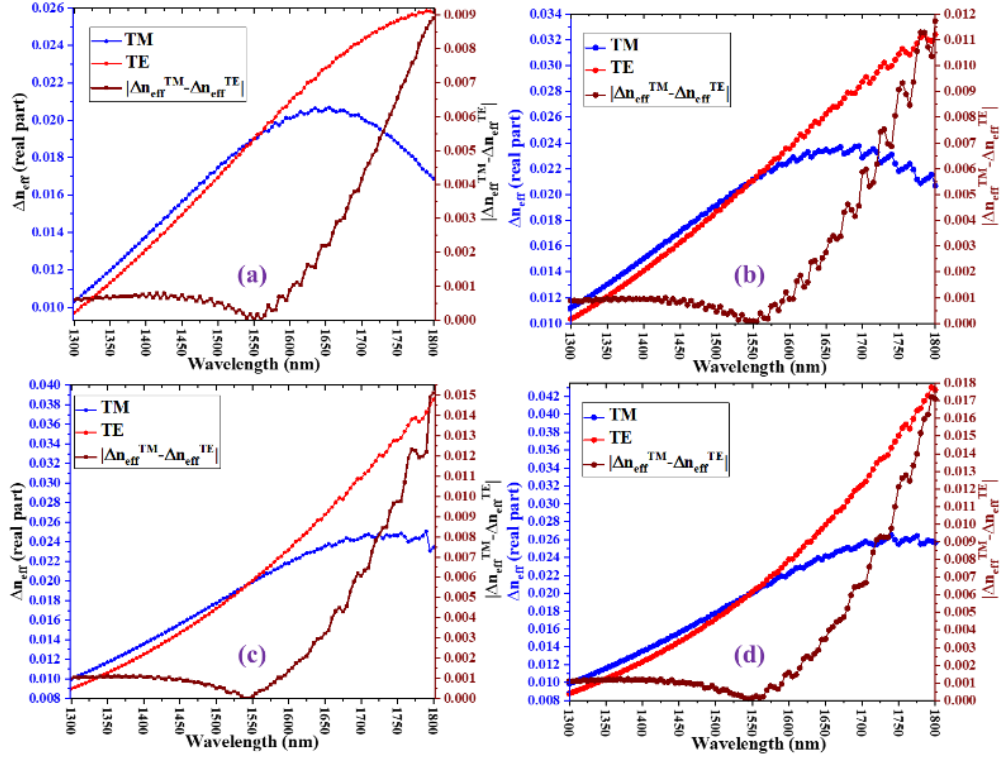

Fig. 13. The real parts of $\Delta n_{eff}^{TM}$, $\Delta n_{eff}^{TE}$, and $|\Delta n_{eff}^{TM} - \Delta n_{eff}^{TE}|$ as the functions of wavelength for the waveguides with optimized $W_{Si}$ at the second intersection point (Table 1) for a) 3-side-graphene-covered buried waveguide b) 4-side-graphene-covered buried waveguide c) 3-side-graphene-covered ridge waveguide and d) 4-side-graphene-covered ridge waveguide

It should be noted that in the above discussion, the curves in Fig. 13 (and those in Fig. 4) are not smooth like their counterparts in Fig. 12 (and Fig. 3) and it can clearly be seen they have some ripples. The physical explanation for these ripples as was mentioned at the end of section 3 is the surface modes on the graphene layers which are excited at high chemical potentials (see Fig. 5).

## 7. Conclusion

In this paper, four rectangular graphene-based Si waveguides in the ridge and buried geometries (with a fixed height) were introduced as polarization-insensitive amplitude and phase modulators. This claim was verified by numerical simulation. Although the proposed structures were originally designed for the single wavelength of 1.55 $\mu m$, the polarization independence of these modulators was conserved in a good wavelength range. Therefore, in addition to the polarization-insensitive feature of these optimized graphene-based Si waveguides as a result of the large values of MPA (nearly 0.3 $dB/\mu m$ at $\lambda = 1.55\ m$) and $\Delta n_{eff}$ (nearly 0.02 at $\lambda = 1550\ nm$), they are highly eligible for small footprint modulators in integrated photonic devices.

### References

[1] Romagnoli M, Sorianello V, Midrio M, Koppens F H L, Huyghebaert C, Neumaier D, Galli P, Templ W, D'Errico A and Ferrari A C 2018 Graphene-based integrated photonics for next-generation datacom and telecom *Nat. Rev. Mater.* **3** 392–414

[2]     Reed G T, Mashanovich G, Gardes F Y and Thomson D J 2010 Silicon optical modulators *Nat. Photonics* **4** 518–26
[3]     Sun Z, Martinez A and Wang F 2016 Optical modulators with 2D layered materials *Nat. Photonics* **10** 227–38
[4]     Kaplan A and Ruschin S 2000 Layout for polarization insensitive modulation in LiNbO3 waveguides *IEEE J. Sel. Top. Quantum Electron.* **6** 83–7
[5]     Li G L and Yu P K L 2003 Optical Intensity Modulators for Digital and Analog Applications *J. Light. Technol.* **21** 2010–30
[6]     Liu M, Yin X, Ulin-Avila E, Geng B, Zentgraf T, Ju L, Wang F and Zhang X 2011 A graphene-based broadband optical modulator *Nature* **474** 64–7
[7]     Liu M, Yin X and Zhang X 2012 Double-layer graphene optical modulator. *Nano Lett.* **12** 1482–5
[8]     Gosciniak J and Tan D T H 2013 Theoretical investigation of graphene-based photonic modulators *Sci. Rep.* **3**
[9]     Ye S, Wang Z, Tang L, Zhang Y, Lu R and Liu Y 2014 Electro-absorption optical modulator using dual-graphene-on-graphene configuration *Opt. Express* **22** 26173
[10]    Ye C, Khan S, Li Z R, Simsek E and Sorger V J 2014 λ-Size ito and graphene-based electro-optic modulators on soi *IEEE J. Sel. Top. Quantum Electron.* **20**
[11]    Dalir H, Xia Y, Wang Y and Zhang X 2016 Athermal Broadband Graphene Optical Modulator with 35 GHz Speed *ACS Photonics* **3** 1564–8
[12]    Hu Y, Pantouvaki M, Van Campenhout J, Brems S, Asselberghs I, Huyghebaert C, Absil P and Van Thourhout D 2016 Broadband 10 Gb/s operation of graphene electro-absorption modulator on silicon *Laser Photonics Rev.* **10** 307–16
[13]    Kleinert M, Herziger F, Reinke P, Zawadzki C, de Felipe D, Brinker W, Bach H-G, Keil N, Maultzsch J and Schell M 2016 Graphene-based electro-absorption modulator integrated in a passive polymer waveguide platform *Opt. Mater. Express* **6** 1800
[14]    Abdollahi Shiramin L and Van Thourhout D 2017 Graphene Modulators and Switches Integrated on Silicon and Silicon Nitride Waveguide *IEEE J. Sel. Top. Quantum Electron.* **23**
[15]    Gosciniak J, Tan D T H and Corbett B 2015 Enhanced performance of graphene-based electro-absorption waveguide modulators by engineered optical modes *J. Phys. D. Appl. Phys.* **48** 235101
[16]    Ralević U, Isić G, Vasić B, Gvozdić D and Gajić R 2015 Role of waveguide geometry in graphene-based electro-absorptive optical modulators *J. Phys. D. Appl. Phys.* **48**
[17]    Li W, Chen B, Meng C, Fang W, Xiao Y, Li X, Hu Z, Xu Y, Tong L, Wang H, Liu W, Bao J and Shen Y R 2014 Ultrafast all-optical graphene modulator *Nano Lett.* **14** 955–9
[18]    Zhang H, Healy N, Shen L, Huang C C, Hewak D W and Peacock A C 2016 Enhanced all-optical modulation in a graphene-coated fibre with low insertion loss *Sci. Rep.* **6** 23512
[19]    Lu Z and Zhao W 2012 Nanoscale electro-optic modulators based on graphene-slot waveguides *J. Opt. Soc. Am. B* **29** 1490–6
[20]    Mohsin M, Neumaier D, Schall D, Otto M, Matheisen C, Giesecke A L, Sagade A a. and Kurz H 2015 Experimental verification of electro-refractive phase modulation in graphene *Sci. Rep.* **5** 10967
[21]    Ye S-W, Yuan F, Zou X-H, Shah M K, Lu R-G and Liu Y 2017 High-Speed Optical Phase Modulator Based on Graphene-Silicon Waveguide *IEEE J. Sel. Top. Quantum Electron.* **23** 1–5
[22]    Sorianello V, Midrio M, Contestabile G, Asselberghs I, Van Campenhout J, Huyghebaert C, Goykhman I, Ott A K, Ferrari A C and Romagnoli M 2018 Graphene-silicon phase modulators with gigahertz bandwidth *Nat. Photonics* **12** 40–4
[23]    Hu X, Zhang Y, Chen D, Xiao X and Yu S 2019 Design and Modeling of High Efficiency Graphene Intensity/Phase Modulator Based on Ultra-Thin Silicon Strip Waveguide *J. Light. Technol.* **37** 2284–92
[24]    Midrio M, Boscolo S, Moresco M, Romagnoli M, De Angelis C, Locatelli A and Capobianco A-D 2012 Graphene–assisted critically–coupled optical ring modulator *Opt. Express* **20** 23144
[25]    Qiu C, Gao W, Vajtai R, Ajayan P, Kono J and Xu Q 2014 Efficient Modulation of 1.55 µm Radiation with Gated Graphene on a Silicon Micro-ring Resonator *Nano Lett.* 1–5
[26]    Phare C T, Daniel Lee Y-H, Cardenas J and Lipson M 2015 Graphene electro-optic modulator with 30 GHz bandwidth *Nat. Photonics* **9** 511–4
[27]    Zhou F and Liang C 2019 The absorption ring modulator based on few-layer graphene *J. Opt. (United Kingdom)* **21**
[28]    Pang C, Lu H, Xu P, Qian H, Liu X, Liu X, Li H and Yang Q 2016 Design of hybrid structure for fast and deep surface plasmon polariton modulation *Opt. Express* **24** 17069
[29]    Radko I P, Bozhevolnyi S I and Grigorenko A N 2016 Maximum modulation of plasmon-guided modes by graphene gating *Opt. Express* **24** 8266
[30]    Qu S, Ma C and Liu H 2017 Tunable graphene-based hybrid plasmonic modulators for subwavelength confinement *Sci. Rep.* **7** 1–8
[31]    Hao R, Jiao J, Peng X, Zhen Z, Dagarbek R, Zou Y and Li E 2019 Experimental demonstration of a graphene-based hybrid plasmonic modulator *Opt. Lett.* **44** 2586
[32]    Chi J, Liu H, Huang N and Wang Z 2019 A broadband enhanced plasmonic modulator based on double-layer graphene at mid-infrared wavelength *J. Phys. D. Appl. Phys.* **52**
[33]    Phatak A, Cheng Z, Qin C and Goda K 2016 Design of electro-optic modulators based on graphene-on-

silicon slot waveguides *Opt. Lett.* **41** 2501
[34]  Kovacevic G, Phare C, Set S Y, Lipson M and Yamashita S 2018 Ultra-high-speed graphene optical modulator design based on tight field confinement in a slot waveguide *Appl. Phys. Express* **11**
[35]  Ji L, Zhang D, Xu Y, Gao Y, Wu C, Wang X, Li Z and Sun X 2019 Design of an Electro-Absorption Modulator Based on Graphene-on-Silicon Slot Waveguide *IEEE Photonics J.* **11** 1–11
[36]  He X, Xu M and Zhang X 2016 Theoretical investigation of a broadband all-optical graphene-microfiber modulator *J. Opt. Soc. Am. B* **33** 2588
[37]  Cheng Z, Tsang H K, Wang X, Chen X, Xu K and Xu J Bin 2013 Polarization dependent loss of graphene-on-silicon waveguides *2013 IEEE Photonics Conf. IPC 2013* **5** 460–1
[38]  Ye S-W, Liang D, Lu R-G, Shah M K, Zou X-H, Yuan F, Yang F and Liu Y 2017 Polarization-Independent Modulator by Partly Tilted Graphene-Induced Electro-Absorption Effect *IEEE Photonics Technol. Lett.* **29** 23–6
[39]  Shah M K, Lu R, Peng D, Ma Y, Ye S, Zhang Y, Zhang Z and Liu Y 2017 Graphene-Assisted Polarization-Insensitive Electro-absorption Optical Modulator *IEEE Trans. Nanotechnol.* **16** 1004–10
[40]  Hu X and Wang J 2018 Design of graphene-based polarization-insensitive optical modulator *Nanophotonics* **7** 651–8
[41]  Zou X, Zhang Y, Li Z, Yang Y, Zhang S, Zhang Z, Zhang Y and Liu Y 2019 Polarization-Insensitive Phase Modulators Based on an Embedded Silicon-Graphene-Silicon Waveguide *Appl. Sci.* **9** 429
[42]  Meng Y, Ye S, Shen Y, Xiao Q, Fu X, Lu R, Liu Y and Gong M 2018 Waveguide engineering of graphene optoelectronics-modulators and polarizers *IEEE Photonics J.* **10**
[43]  Yang Z, Lu R, Cai S, Wang Y, Zhang Y, Wang X and Liu Y 2019 A CMOS-compatible and polarization-insensitive graphene optical modulator *Opt. Commun.* **450** 130–5
[44]  Xu Y, Li F, Kang Z, Huang D, Zhang X, Tam H-Y and Wai P 2019 Hybrid Graphene-Silicon Based Polarization-Insensitive Electro-Absorption Modulator with High-Modulation Efficiency and Ultra-Broad Bandwidth *Nanomaterials* **9** 157
[45]  Geim A K and Novoselov K S 2009 The rise of graphene *Nanosci. Technol. A Collect. Rev. from Nat. Journals* **6** 11–9
[46]  Amin R, Ma Z, Maiti R, Khan S, Khurgin J B, Dalir H and Sorger V J 2018 Attojoule-efficient graphene optical modulators *Appl. Opt.* **57** D130
[47]  Kim J, Son H, Cho D J, Geng B, Regan W, Shi S, Kim K, Zettl A, Shen Y R and Wang F 2012 Electrical control of optical plasmon resonance with graphene *Nano Lett.* **12** 5598–602
[48]  Bao Q, Zhang H, Wang B, Ni Z, Lim C H Y X, Wang Y, Tang D Y and Loh K P 2011 Broadband graphene polarizer *Nat. Photonics* **5** 411–5